\documentclass{aa}
\usepackage{graphics}

\newcommand{\apj}{ApJ}

\newcommand{\aap}{A\&A}

\begin{document}

\thesaurus{03(03.09.2; 03.13.4)}
\title{Interferometric array design: \\
optimizing the locations of the antenna pads}
\author{Fr\'ed\'eric Boone}
\institute{ frederic.boone@obspm.fr\\
DEMIRM, Observatoire de Paris,
61 av. de l'Observatoire,
F-75014 Paris}
\date{Received 28 June 2001 / Accepted }
\authorrunning{F.Boone}
\titlerunning{Interferometric array design I}
\maketitle
\begin{abstract}
The design of an interferometric array should allow optimal instrumental response regarding all possible source positions, times of integration and scientific goals. It should also take into account constraints such as forbidden regions on the ground due to impracticable topography. The complexity of the problem requires one to proceed by steps. A possible approach is to first consider a single observation and a single scientific purpose. A new algorithm is introduced to solve efficiently this particular problem called the configuration problem. It is based on the computation of pressure forces related to the discrepancies between the model (as determined by the scientific purpose) and the actual distribution of Fourier samples. The flexibility and rapidity of the method are well adapted to the full array design. A software named APO that can be used for the design of new generation interferometers such as ALMA and ATA has been developed.
\keywords{Instrumentation: interferometers -- Method: numerical}
\end{abstract}
\section{Introduction}\label{sec:intro}

\begin{figure}

         \resizebox{8.7cm}{!}{\rotatebox{90}{\includegraphics{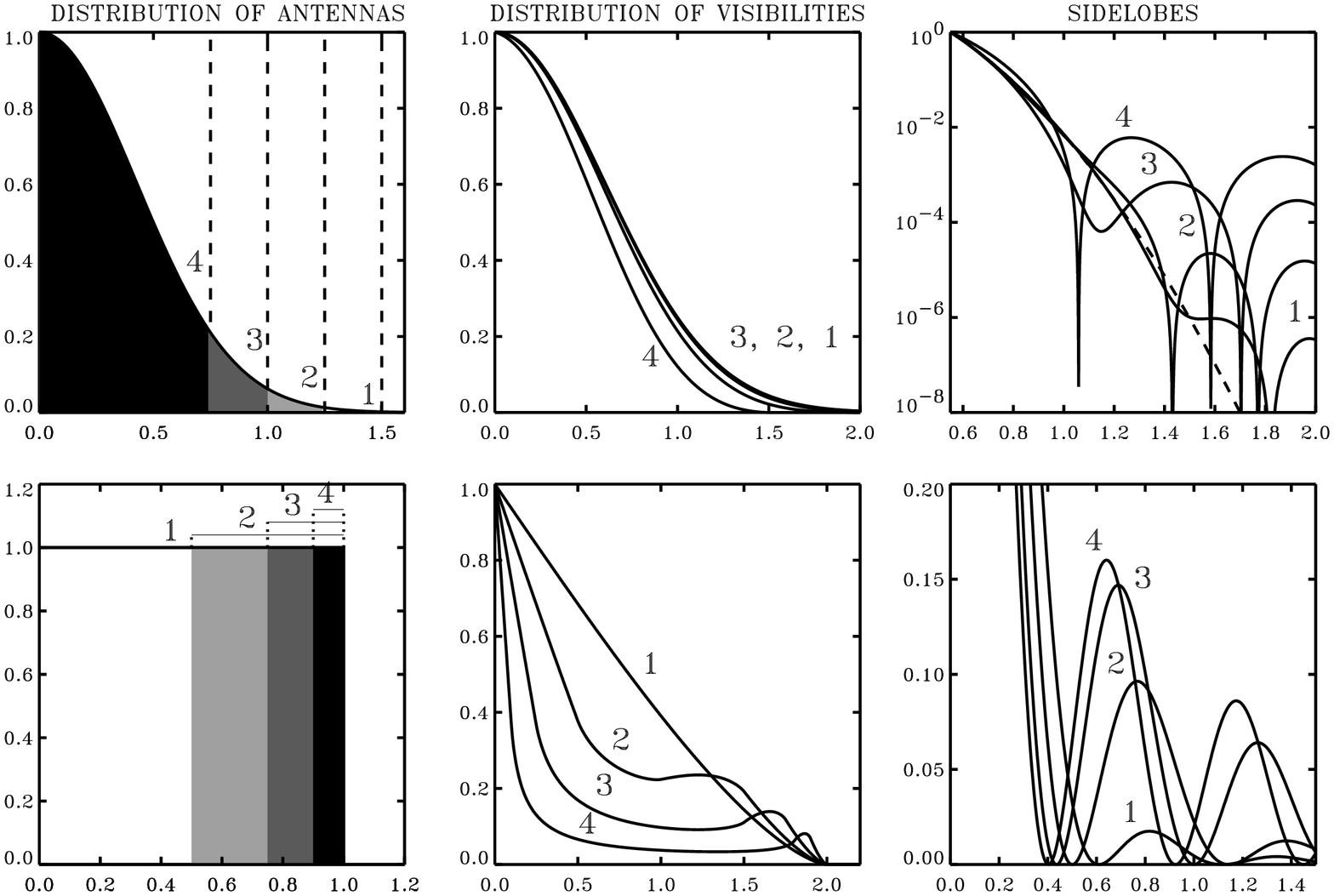}}}
         \resizebox{8.7cm}{!}{\rotatebox{90}{\includegraphics{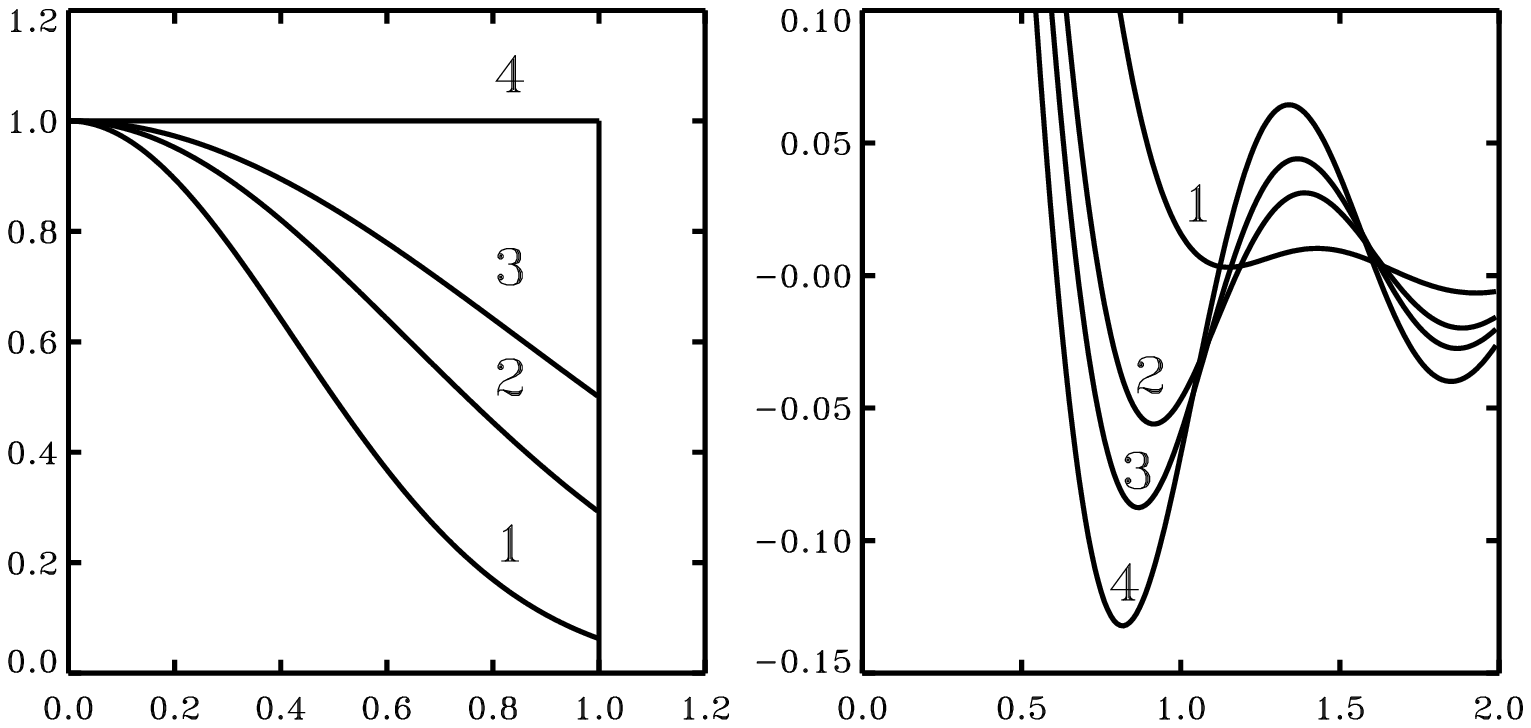}}}

     {\caption[]{ Illustration of the relationship between distribution of antennas, ${\cal A}$, distribution of Fourier samples, ${\cal D}$, and synthesized beam ${\cal S}$: ${\cal S}={\cal F }\{{\cal D }\}$ and for zenithal snapshot observations: ${\cal D}={\cal A}\otimes{\cal A}$. On the first row Gaussian distributions of antennas truncated at $1.5\times$FWHM, i.e. 27dB (1),  $1.25\times$FWHM, i.e. 19 dB (2), $1\times$FWHM, i.e. 12 dB (3) and $0.75\times$FWHM, i.e. 6.8 dB (4); on the second row a ring distribution of antennas of external radius $R$ and of width $R$ (i.e. a filled disc)(1),  R/2 (2), R/4 (3) and  R/10 (4). The third row shows a family of distributions of Fourier samples discussed in Paper II. They are truncated Gaussians with FWHM$/R=$1, i.e. a truncation level $\Gamma$=12 dB (1), FWHM$/R=$1.5, i.e. $\Gamma$=5 dB (2), FWHM$/R=$2, i.e. $\Gamma$=3 dB (3) and FWHM$/R=\infty$, i.e. $\Gamma$=0 dB (4), $R $ being the truncation radius. It can be shown that such distributions do not admit any solutions for the distribution of antennas. All curves are normalized to be 1 at maximum.}\label{fig:distri}}%
\end{figure}
\begin{figure*}
        \vspace{0cm}

         \resizebox{17cm}{!}{\includegraphics{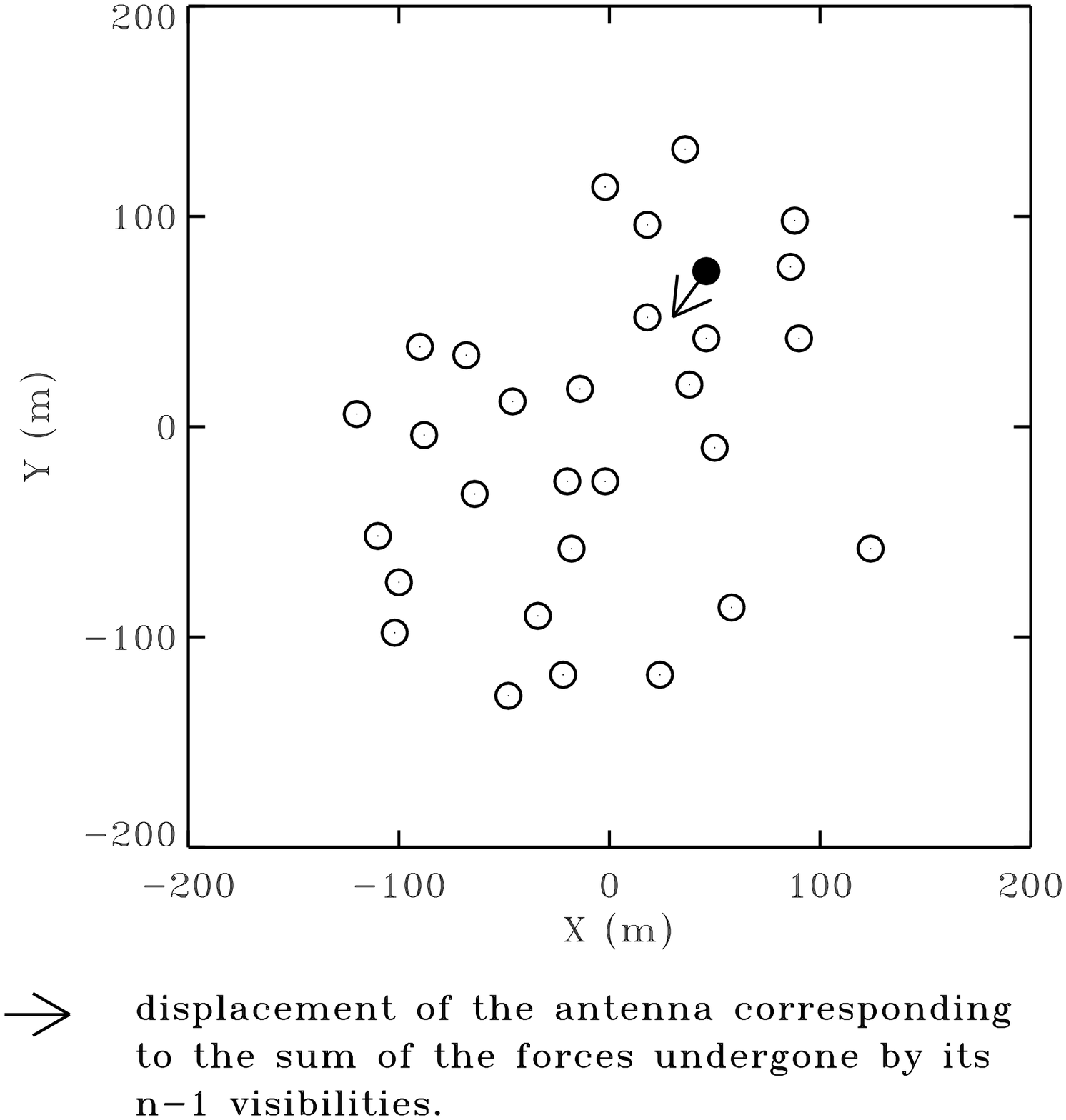}\hspace{0cm}
                    \includegraphics{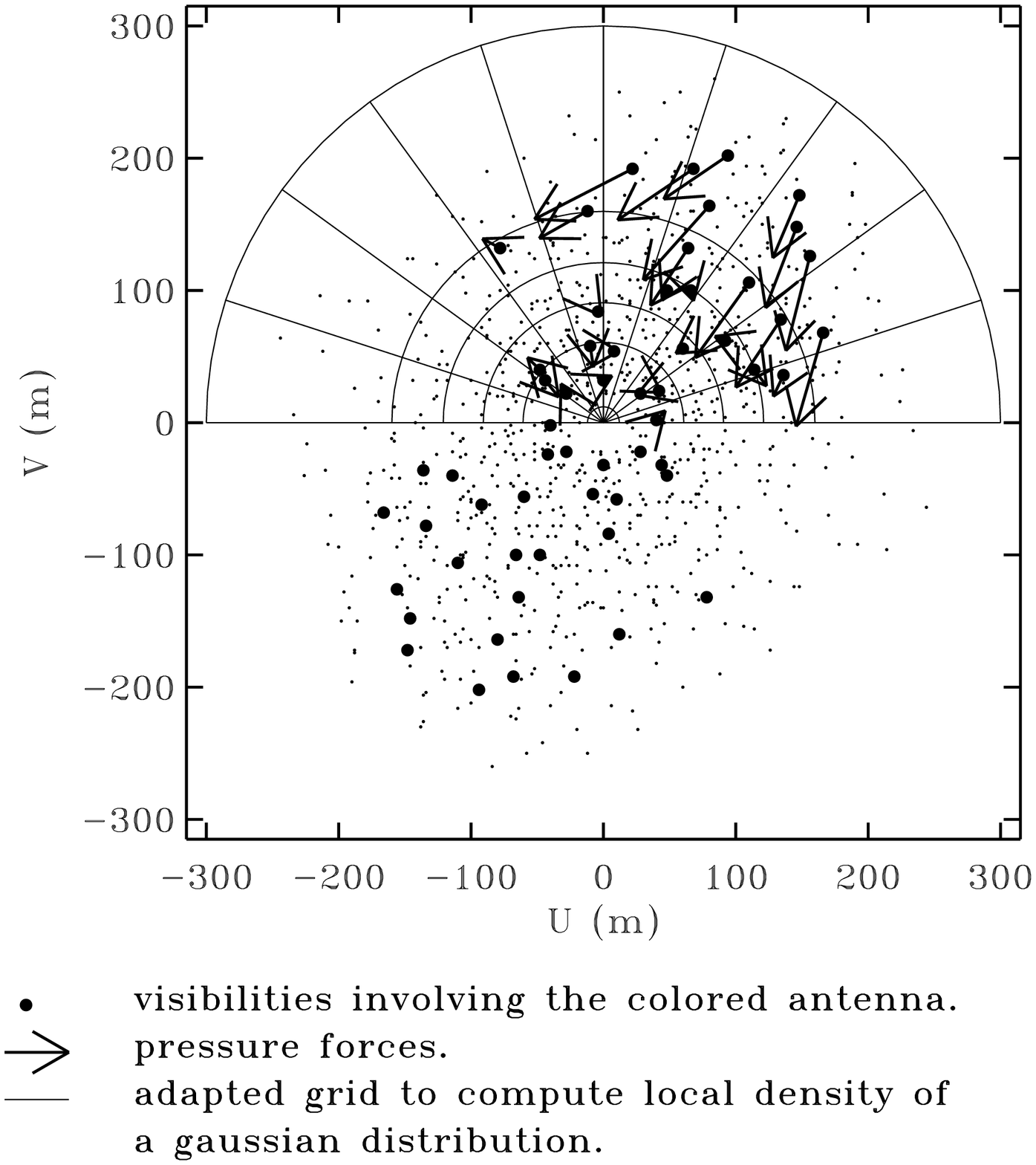}}

     {\caption[]{Illustration of the method: on the right-hand side the antenna plane, on the left-hand side the Fourier plane with the Fourier samples as measured by the array. The density of Fourier samples is computed in the grid represented and a pressure force is derived for each visibility involving the colored antenna. The displacement for the colored antenna is proportional to the average of the pressure forces.}\label{fig:method}}%
\end{figure*}
Designing an interferometric array consists mainly in choosing the locations of the stations (or pads) that will carry the antennas during the observations. In general  more stations than antennas are planned to allow several configurations. An ideal design should ensure optimal configurations regarding all possible observation situations (source positions and durations of observation), scientific purposes (single field imaging, mosaicing, astrometry, detection, ...) and constraints (cost, ground composition and practicability, operation of the instrument, ...). The large number of parameters and sometimes incompatible specifications make this optimization problem complex and difficult to solve globally. The development of radio-interferometric instrumentation has given rise to several publications contributing to some aspects of the problem \nocite{1991QB479.3.T47.5...,cornwell.m38,1993A&A...271..697C,1997ApJ...475..843K,conway.m283,conway.m292,kogan.m171,woody.m270}(e.g., {Thompson} {et~al.} 1991; {Cornwell} 1986; {Cornwell} {et~al.} 1993; {Keto} 1997; {Conway} 2000a,b; {Kogan} 1997; Woody 1999). This paper concentrates on the ``configuration problem'' stated below which can be seen as a first step in the optimization process. A method able to solve this problem is proposed and guidelines on the way to use it for a full array design are presented.

The ``configuration problem'' may be stated as follows: given,
\begin{itemize}
  \item an instrument made of $n_a$ antennas of diameter $D$ and $n_{conf}$ configurations,
  \item a site at a given latitude and with terrain constraints like forbidden regions for the antennas,
  \item an observational situation defined by the position of the source and the duration of the observation,
  \item a model distribution for the Fourier samples as required by the scientific goal,
\end{itemize}
what are the optimal locations for the $n_a$ antennas in the different configurations?

This problem differs from the general ``design problem'' since only a single observation situation and a single scientific purpose are considered. But, as will be shown below, getting over this first obstacle makes the full array design accessible. The relationship between the scientific purpose and the distribution of Fourier samples is central and deserves a complete analysis. This constitutes the subject of a second paper \nocite{booneII}({Boone} 2001b, hereafter Paper II). 

An introduction and a very clear description of the configuration problem is given in \nocite{1997ApJ...475..843K}{Keto} (1997). I shall only recall that for a zenithal snapshot observation the sampling function in Fourier plane (the function composed of Dirac $\delta$-functions at the sample coordinates) is equal to the autocorrelation of the configuration function (the function composed of Dirac $\delta$-functions at the antenna coordinates) without the central point  when the antennas are not correlated with themselves. Several examples are illustrated in Fig.\ref{fig:distri} where the distribution of Fourier samples and the corresponding synthesized beam are given for some continuous two dimensional distributions of antennas. There is generally no  distribution of antennas able to yield a distribution of samples equal to a given 2d-function, since this given function  is not necessarily an autocorrelation. For example it can be shown that there is no solution for a uniform distribution: no 2d  real positive function can yield a top hat function by autocorrelation. More generally an autocorrelation function is necessarily derivable and the distributions represented on the third row of Fig.\ref{fig:distri} do not admit any solution for the distribution of antennas. But their properties, discussed in Paper II, are interesting and it might be worth deriving configurations yielding distributions as close as possible to those ones. Thus, ``solving'' the configuration problem does not mean inverting an autocorrelation product but rather finding the configuration yielding the distribution of samples closest to the target one (it is an inverse problem). The use of an optimization method derives naturally from this observation.

This paper is organized as follows: in Sect.\ref{sec:method} the existing methods are briefly recalled and a new one based on pressure forces is introduced. Sect.\ref{sec:implem} describes how the optimization convergence can be improved and the different kinds of observation (synthesis, multi-configuration, mosaicing) integrated in the implementation. In Sect.\ref{sec:applic} the application of the method to various situations and guidelines for the full array design are presented. Sect.\ref{sec:conclu} gives the conclusion. 


\section{ An optimization driven by pressure forces}\label{sec:method}

\begin{figure*}
        \vspace{0.2cm}
 \resizebox{17cm}{!}{\rotatebox{90}{\includegraphics{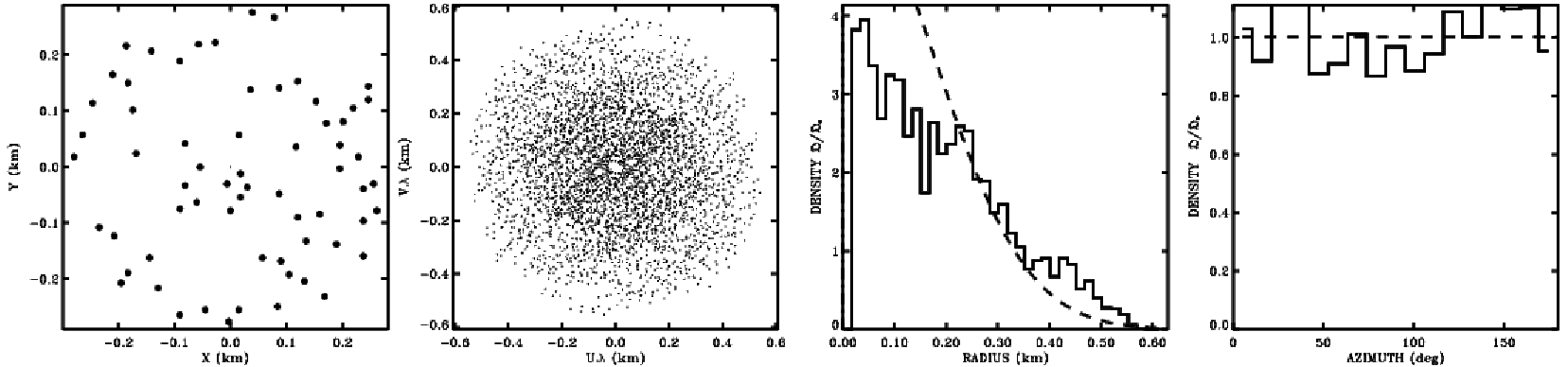}}}
\resizebox{17cm}{!}{\rotatebox{90}{\includegraphics{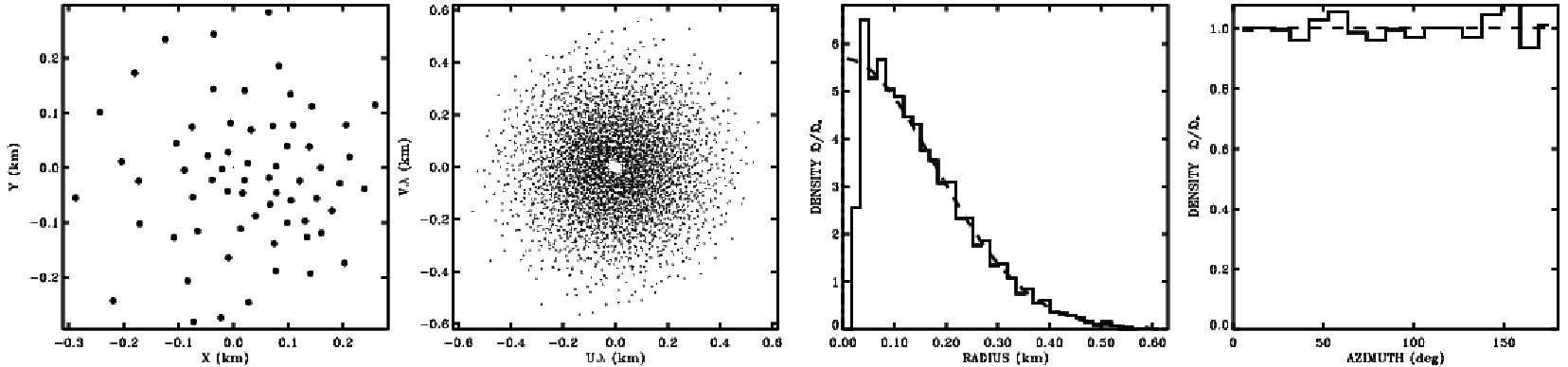}}}

     {\caption[]{Example of optimization: a configuration of 64 antennas is optimized for a Gaussian distribution of Fourier samples. The first row shows the initial randomly picked configuration, its Fourier samples and the radial and azimuthal profiles of the density of Fourier samples. The second row shows the same for the optimized configuration. The model distribution is represented by the dashed lines in the radial and azimuthal profiles. The optimal distribution of antennas is expected to be Gaussian, this property is checked in Fig.\ref{fig:conv}.}\label{fig:gaussian}}%

        \vspace{0.2cm}

         \resizebox{17cm}{!}{\includegraphics{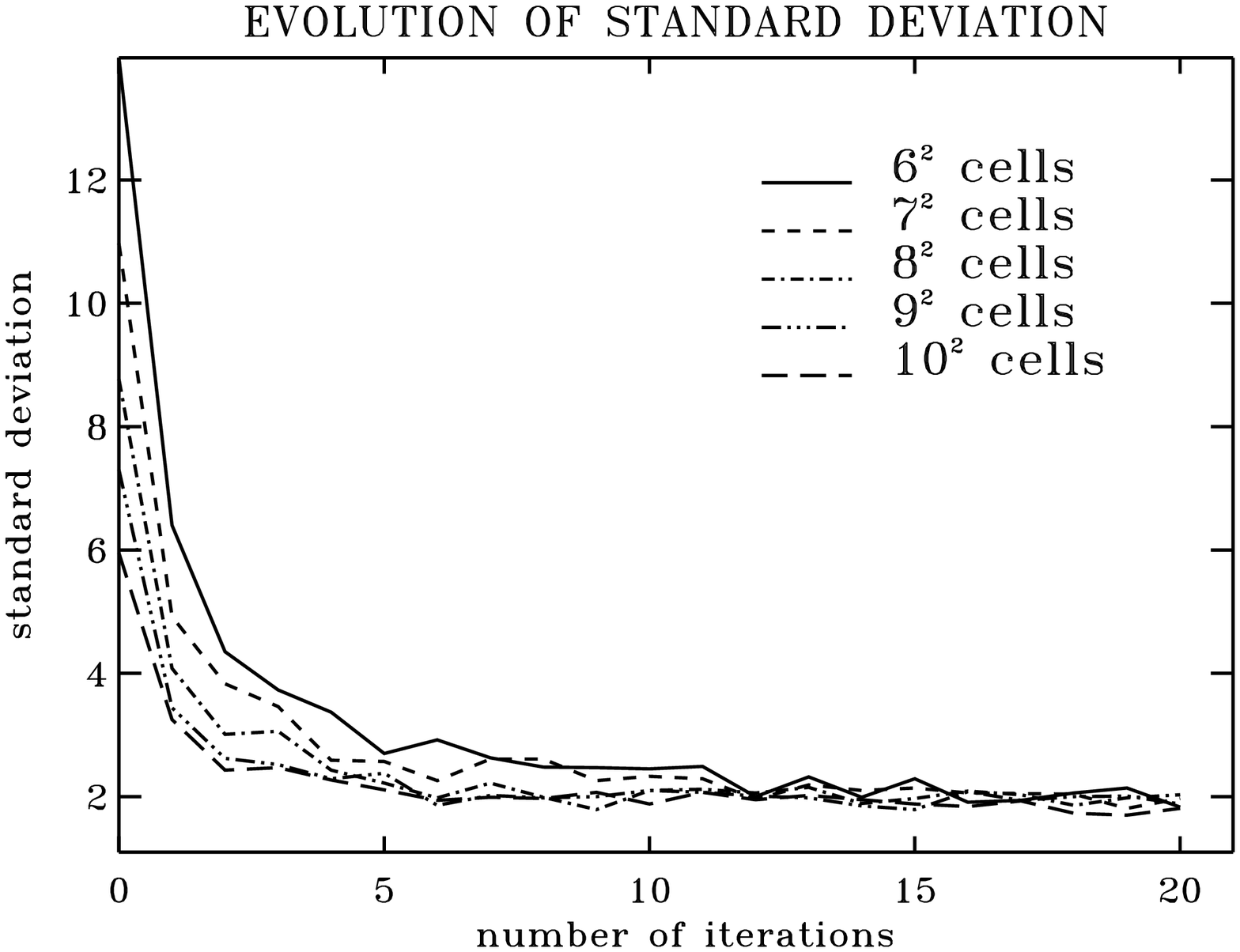}
                    \hspace{0cm}
                    \includegraphics{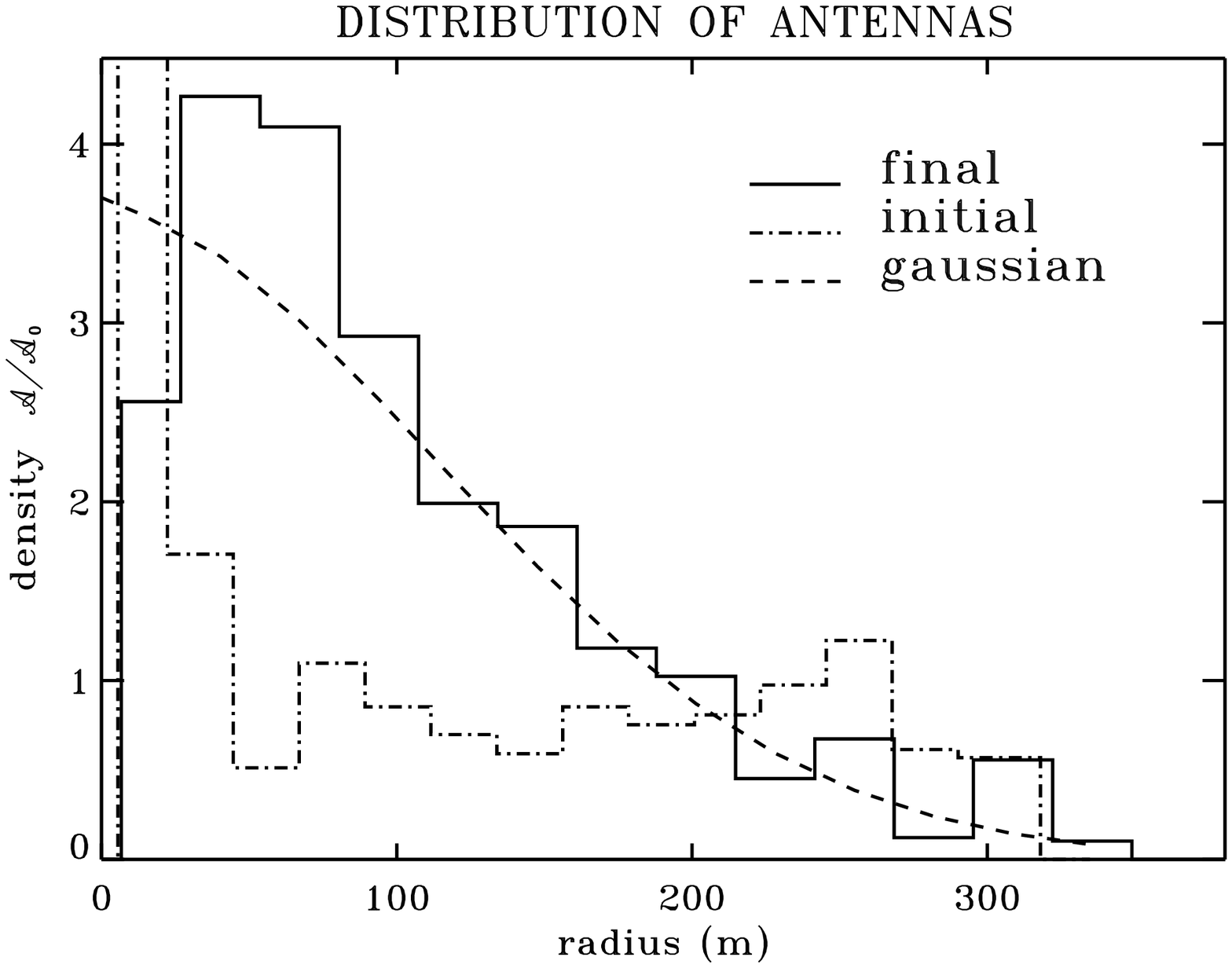}}
     {\caption[]{Convergence and quality of the result for the optimization illustrated in Fig.\ref{fig:gaussian}. On the left-hand side the variations of the standard deviation at various resolutions (for grids made of $6^2$ to $10^2$ cells per quadrant) show the rapid and stable convergence of the method. Each iteration lasts about 10s in time. On the right-hand side the distribution of antennas from the optimized configuration fits quite well a Gaussian distribution. This is indeed the result expected for a Gaussian distribution of Fourier samples.}\label{fig:conv}}%

\end{figure*}

Mainly two methods are currently used by the community to solve the configuration problem: one was developed by \nocite{1997ApJ...475..843K}{Keto} (1997) in the frame of the SMA project and the other by \nocite{kogan.m171}{Kogan} (1997) in the frame of the ALMA project.

In the first method, randomly picked positions in the uv-plane
pull the nearest Fourier samples by moving the corresponding antennas in the same direction. The random positions are picked with a probability distribution equal to the final wanted distribution of Fourier samples. With a large number of iterations, and thanks to a neural network, the algorithm converges to a solution.  Although this method allows one to get some interesting results it is still computationally  expensive and can only handle a small number of antennas for snapshot observations only. It was argued in \nocite{1997ApJ...475..843K}{Keto} (1997) that the effect of Earth rotation synthesis on the instrumental response could be easily derived and compensated as the coordinates of the samples result from rotations and projections of a zenithal snapshot observation. It was also suggested to optimize an array for zenithal observations only, considering it as the average source position. The point of view supported here is quite different. First, as will be shown in Sect.\ref{sec:applic} with some examples, the dependency of the configurations on the source position for long track observations is obvious and its prediction from a zenithal snapshot observation is not trivial. Second, sources at low elevations require a substantial elongation of the array in the north-south direction (as well as a rearrangement of the antennas different for the northern and southern sources). Therefore, if several configurations are possible (i.e. the number of pads is greater than the number of antennas) it is better to optimize at least 3 different configurations corresponding to  zenithal, northern and southern observations. For that purpose it is necessary to  include the earth rotation synthesis in the algorithm. 

Another approach of the problem was proposed in \nocite{kogan.m171}{Kogan} (1997). From the relationship relating the synthesized beam to the antenna positions, an analytical solution for the displacements of the antennas lowering the side lobe level at a given position on the synthesized beam was derived. If this method can improve a configuration from the side lobe level  point of view it is not able to find an optimal configuration for any arbitrary distribution of Fourier samples and thus is not able to answer the configuration problem in the general sense. Furthermore, it is shown in Paper II that even for imaging purposes the target distribution of samples should not always be the one that minimizes side lobes. If this method can be useful in some particular cases (e.g. when it is wanted to optimize the imaging quality of a predefined configuration shape known to provide the required sampling for snapshot observations), it is argued the configuration problem should be treated exclusively from the Fourier plane point of view, i.e. for a wanted distribution of Fourier samples. Again, what an interferometer actually measures is a set of angular spatial frequencies and to each scientific goal an optimal distribution of samples should be defined. This distribution may not necessarily yield low side lobes but should allow to recover the relevant information with the highest possible sensitivity (see Paper II).

The method proposed here is based on the fact that while it is usually not possible to get a direct solution for the antenna positions it is easy for any configuration to figure out how the Fourier samples should be moved to ``improve'' the distribution, or make it more similar to the model. For instance, if there is a hole in the distribution where it should not be, some of the nearby Fourier samples should be moved to fill-in this hole. The question is then: how should the antennas move in order to allow such an improvement?

There is a direct geometrical relationship between each sample  coordinates and the positions of two antennas, but moving one antenna implies moving $n_a-1$ Fourier samples and it would not necessarily be an improvement to move one antenna according to only one sample. The approach suggested here  is to move each antenna according to the $n_a-1$ Fourier samples in which it is involved, i.e. according to the average of the $n_a-1$ displacements these Fourier samples should undergo to improve the distribution.
 
The way a sample should be moved is numerically derived by computing the local ``excess-density'' gradient, $\vec{G}$. By excess-density is meant the difference between the actual density, ${\cal D}(u,v)$ and the model density, ${\cal D}_m(u,v)$:
\begin{equation}
  \vec{G}(u,v)=\vec{\nabla} ({\cal D}(u,v) -{\cal D}_m(u,v))
\end{equation}
The sample of coordinate $(u,v)$ should move in the opposite direction of this gradient vector by an amount proportional to its amplitude. The vector $-\vec{G}$ may be interpreted as a pressure force undergone by the Fourier samples, either pulled out overcrowded regions, or sucked into insufficiently covered regions.

The method is illustrated in Fig.\ref{fig:method}. For each antenna the $n_a-1$ gradient vectors corresponding to the $n_a-1$ Fourier samples are computed, transformed according to the geometrical transformation relating the Fourier plane to the ground plane and the antenna moved according to the average of these vectors. The displacement $\vec{D}$ for one antenna is given by:
\begin{equation}\label{eq:D}
\vec{D}=g \sum_{i=1}^{n_a-1} \vec{M} \vec{G}(u_i,v_i)
\end{equation}
where $g$ is an {\it ad hoc} gain factor and $\vec{M}$ is the transformation matrix from the uv-plane to the ground plane:
\begin{equation}
\vec{M}=\left(
\begin{array}{cc}                                
\frac{\sin(\delta)\sin(\lambda)\cos(H)+\cos(\delta)\cos(\lambda)}{\cos(\delta-\lambda)} & \frac{\sin(H)\sin(\lambda)}{\cos(\delta-\lambda)} \\
-\frac{\sin(\delta)\sin(H)}{\cos(\delta-\lambda)} & \frac{\cos(H)}{\cos(\delta-\lambda)} \\
\end{array}
\right)
 \end{equation}
with $\delta$ the source declination, $\lambda$ the site latitude and $H$ the hour angle. 

By repeating this operation for each antenna and iterating, the configuration should converge to an optimal solution. Ideally, once the optimal configuration is reached, the forces should equal zero everywhere  meaning that the distribution obtained is equal to the model distribution. In practice, as the model distribution is not necessarily an autocorrelation function it may not be  possible to exactly fit it with the distribution of Fourier samples. In addition, a continuous distribution can not be perfectly fitted by the density of a limited number of points. Hence, the forces of the final configuration are never null. 

Without going into a full convergence analysis, it seems reasonable to think that, if there is still some discrepancies between the model and the actual distribution, at least for one antenna the sum of the pressure forces on its Fourier samples will not cancel out and it will be moved to improve (if it can be improved) the distribution. If the distribution can not be improved the configuration will oscillate around the optimal one and by decreasing $g$ along iterations it should converge. In other words the method seems capable of avoiding local minima and it was therefore decided to stop the optimization process when the standard deviation between the actual distribution and the model reaches a stable minimum. If there is no rigorous proof that this  configuration is indeed the best one, one can hope that it is close to it and at least that it fits satisfactorily the requirements on the distribution of Fourier samples.

This method offers several advantages. First, the convergence is self-driven: there is no need to optimize blindly any quantity in a space of parameters nor to use random picking \nocite{1997ApJ...475..843K}(like in {Keto} 1997). At each step the distribution contains in itself the way the configuration should evolve. The risk of getting caught in resonant oscillations is avoided by allowing only small displacements for the antennas i.e. by setting $g$ to a small value. Second, the main computational cost is in the calculation of the pressure force which is a simple operation. Finally, it is very flexible and may be used for any 2d model distribution, with ground constraints, and can handle Earth rotation synthesis, multi-configuration observations and mosaicing.

Incidentally, it can be noted that this method is a variant of the {\it steepest descent method}. Indeed, $\vec{D}$ as given in Eq.\ref{eq:D} represents the local downhill gradient of the integral of the excess-density.
\section{Implementation}\label{sec:implem}

The execution speed and the convergence are highly dependent on the algorithm used for the computation of the density gradient. The simplest way to calculate this gradient is to count the number of Fourier samples in each cell of a grid, and then, the difference between adjacent cells. The choice of the grid is crucial for the convergence of the method. Consider the case of a Gaussian model distribution: if the grid is orthogonal many cells in the outer part of the distribution will be empty causing gradients with the neighbors which may contain only one sample. In other words the program will be sensible to gradients arising from the discrete nature of the sampling function. In addition it will give the same weight to a gradient over a region containing lot of samples in the center and a similar gradient concerning only a few samples on the outer part. Finally,  the circular boundary will cross the square cells and the area of each of these cells being outside the boundary will depend upon the coordinates of the cell. The distribution at the boundaries will consequently be out of control. 

For these reasons it is optimal to use an adaptive circular grid. That is, a circular grid for which the number of Fourier samples in each cell is constant when the distribution is  equal to the model. Thus, in the Gaussian model example where the model density is given by:
\begin{equation}
{\cal D}_m(R)=C_1 \exp\!\left(-C_2\, r^2\right)
\end{equation}
if the angular size of the cells is $\Delta \psi$ and the inner boundary of a cell is at radius $R_n$ then, for this cell to contain $N$ Fourier samples, the outer boundary must be at radius:
\begin{equation}
R_{n+1}=\sqrt{\frac{1}{C_2}\ln \!\left(\exp\!\left(-C_2 \,{R_n}^2\right)-\frac{2 \,N\, C_2}{\Delta\psi\, C_1}\right)}
\end{equation}
 Notice that since the distribution is centrally symmetric, the grid has to cover only half of the sampled uv-disc (see Fig.\ref{fig:method}).

However, optimizing a configuration with only one grid do not ensure the resulting distribution of samples to fit the model at all resolutions. It can show strong defects at larger or smaller scales than the average cell size of the grid. In order to optimize the distribution at all resolutions simultaneously several grids have to be used simultaneously. For example to optimize a 64-antennas array it was found optimal to use 7 grids of sizes $6^2$ up to $13^2$ cells per quadrant.

For topography constraints the simplest situation has been considered: a digital mask was used to define forbidden regions for the antennas. When moving an antenna, if the destination falls into such a forbidden area the antenna  is placed either before or after the area depending on which is the nearest to the original destination.  More complex constraints may also be considered, e.g. in the form of pressure forces on the antennas arising from the local level of forbidding. 

In the case of earth rotation synthesis the geometrical transformation from uv-plane to antenna plane, $\vec{M}$, is different for each sample of a given antenna pair, in addition different weights are given to the Fourier samples according to the elevation of the source, i.e. to the level of noise. The choice of the averaging time separating two measurements is a compromise between computing time and good sampling of the largest baseline track. It is not related to the real operation of the instrument. For example it can be taken equal to  half an hour for a 8h-observation: 16 points per track might be enough in the sense that taking more points would not change the resulting configuration.

Mosaicing \nocite{1988A&A...202..316C, 1993A&A...271..697C}(see e.g., {Cornwell} 1988; {Cornwell} {et~al.} 1993) can be easily integrated in the program by allowing only segments of the tracks to be sampled. The unsampled parts correspond to the time spent on the other pointings of the mosaic. It is stressed however that the way the tracks are sampled, regularly or by segments in the case of mosaicing, has little impact on the configurations provided that the time spent on the other fields is not too long, i.e. the mosaic is not too large. As shown Sect.\ref{sec:applic} the curvature of the tracks described by the baselines has a strong impact as it determines the way they can be packed to make the density fit as well as possible  the model distribution. This  curvature is affected only when the separation between sampled segments is large i.e. greater than $\frac{1}{2}$h.  Hence, if 1 min is spent on each pointing, the mosaic has to be larger than 30 fields to have an impact on the configuration.

To handle multi-configuration observations the density is computed by adding the Fourier samples measured by all the configurations. Some antennas might take part in several configurations allowing for the situation in which only some of the antennas are moved before observing the same source again. For each configuration uv-plane constraints are given in the form of a maximum and minimum radius for the uv-ring to be sampled. The minimum radius constraint allows to exclude any shadowing between the antennas: shadowing happens when some Fourier samples are inside a radius equal to the projected antenna diameter.

 A C++ library, named APO (Antenna Positions Optimization), was written to optimize the positions of the antennas of any instrument in any of those situations. The flexibility of the object oriented language is well adapted to the implementation of the method. Five main classes were defined corresponding to the description of an instrument, a site, an observational situation, a model distribution of samples and a grid. Each of these classes can be instantiated several times and the optimization can run on all objects. For instance it is possible to: create 7 different grids as mentioned above and run the optimization considering all of them; create several observational situations with an eventual weight for each of them and optimize the instrument accordingly; optimize several instruments on different sites in order to make them complementary regarding some observational situations and scientific purposes (i.e. distribution of samples).

\section{Applications}\label{sec:applic}
\begin{figure*}
\resizebox{17cm}{!}{\rotatebox{90}{\includegraphics{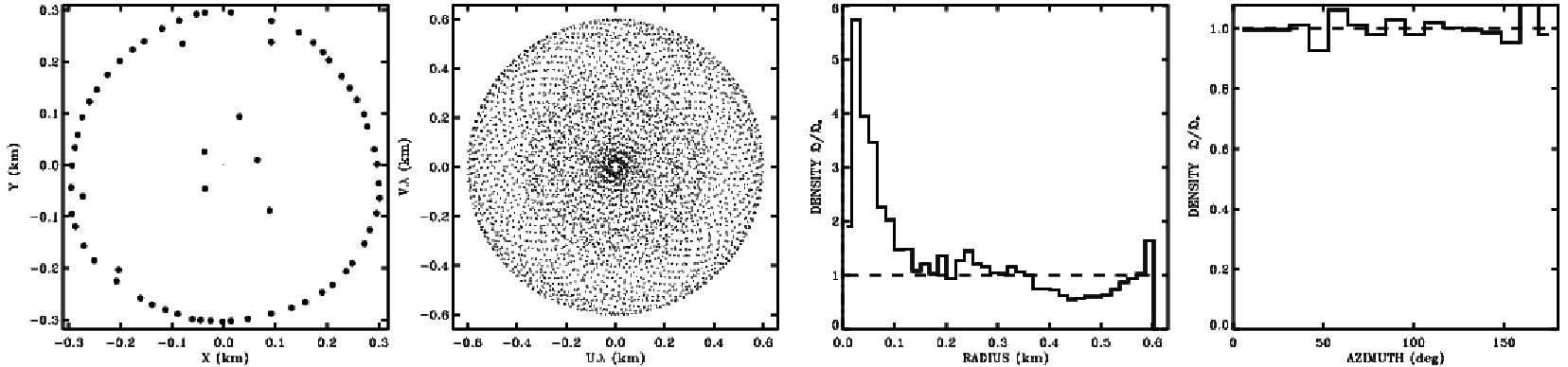}}}
\resizebox{17cm}{!}{\rotatebox{90}{\includegraphics{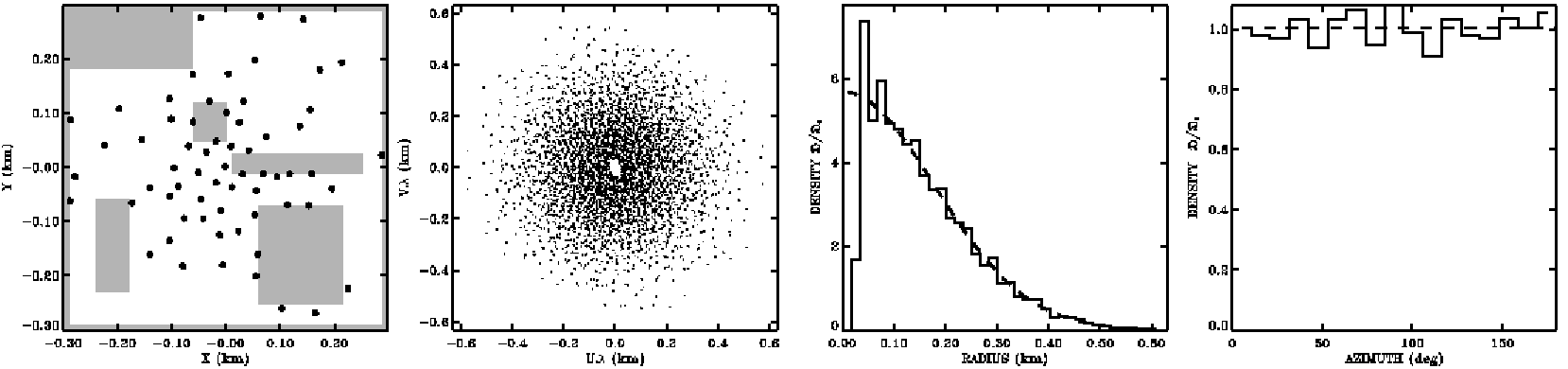}}}
\resizebox{17cm}{!}{\rotatebox{90}{\includegraphics{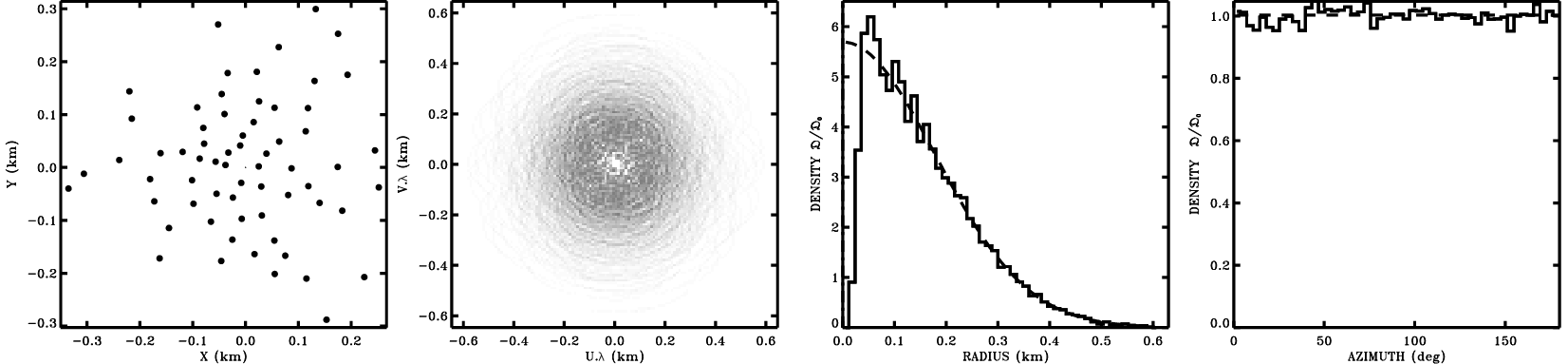}}}
     {\caption[]{Examples of application. On the first row: optimization for a zenithal snapshot observation and a uniform model  distribution of samples (dashed line in the profiles). Second row: same optimization as in Fig.\ref{fig:gaussian} but with ground constraints: the forbidden regions are symbolized by the dark areas. Third row: optimization for the same Gaussian model distribution but for a 6h-observation, 3h on either side of the zenith.}\label{applic}}%
\end{figure*}
\begin{figure*}
\resizebox{17cm}{!}{\rotatebox{90}{\includegraphics{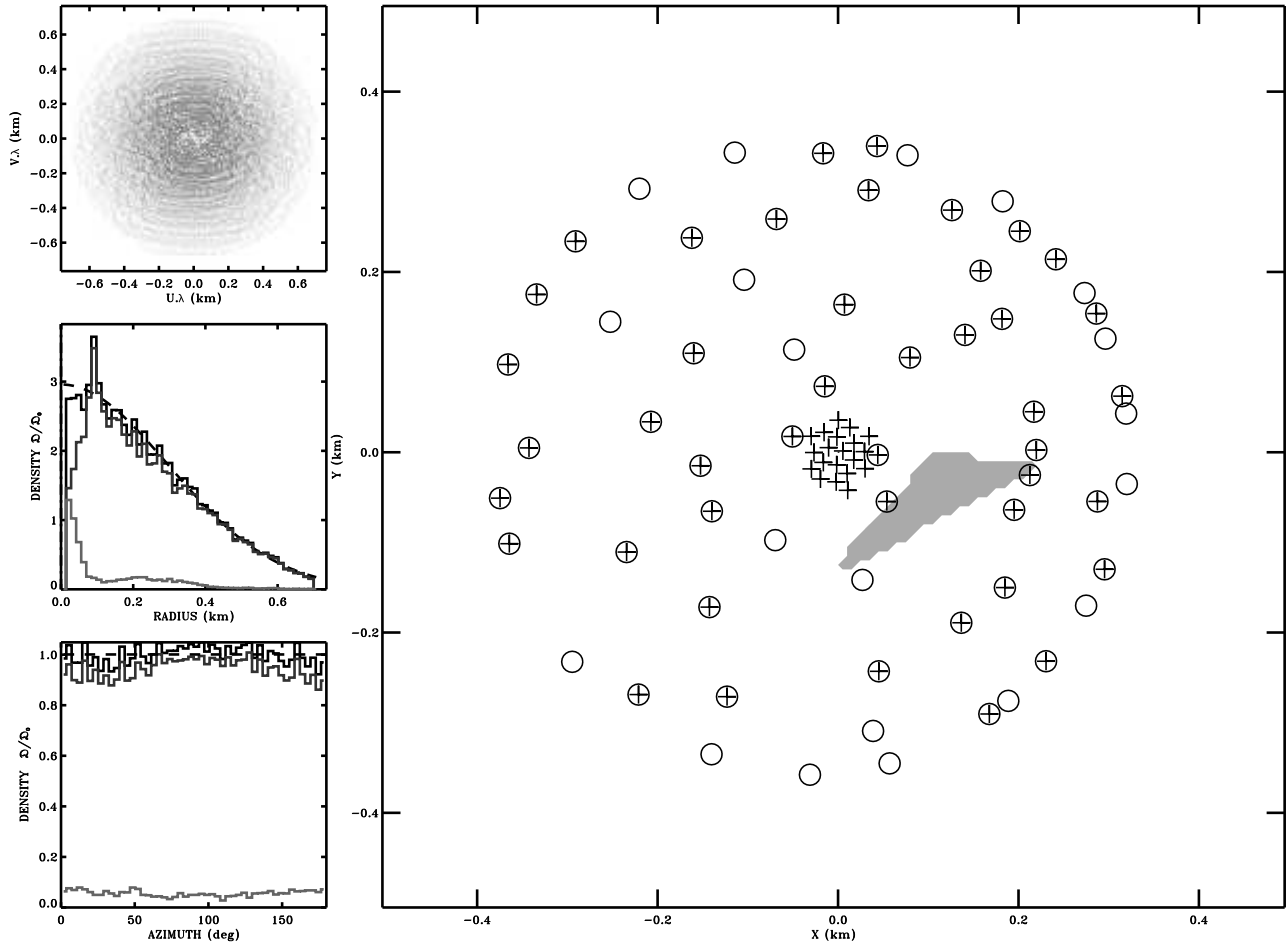}}}

     {\caption[]{Optimization for a Gaussian distribution of samples of 84 pad positions for a 64-antennas array at $-23\deg$ of latitude observing the same source at $-23\deg$ of declination with two configurations having 44 pads in common: a snapshot observation with the antennas at the positions symbolized by the crosses and  a 6h observation with the antennas at the positions symbolized by the circles. The positions symbolized by a cross in a circle correspond to pads involved in both observations. The density of Fourier samples is represented on the top left view graph, as well as its integrated radial and azimuthal profiles in black line in the middle and bottom viewgraphes. The integrated profiles of both configurations are also plotted: in light grey for the snapshot and in dark grey for the long track observation. It can be noticed that the snapshot observation with some antennas packed in the center allows to improve the sensitivity at short spacings.}\label{multiconf}}%
\end{figure*}
\begin{figure*}
\resizebox{17cm}{!}{\rotatebox{90}{\includegraphics{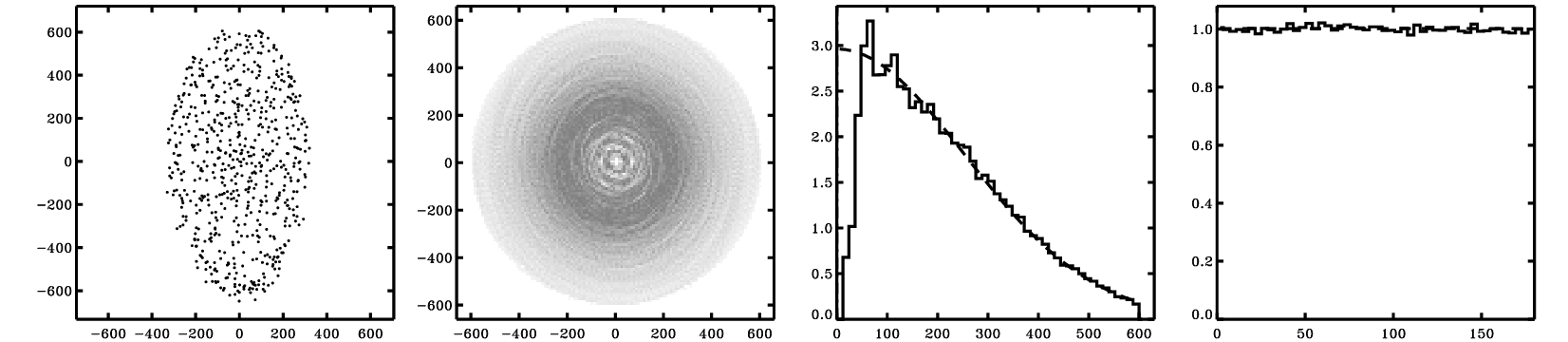}}}
\resizebox{17cm}{!}{\rotatebox{90}{\includegraphics{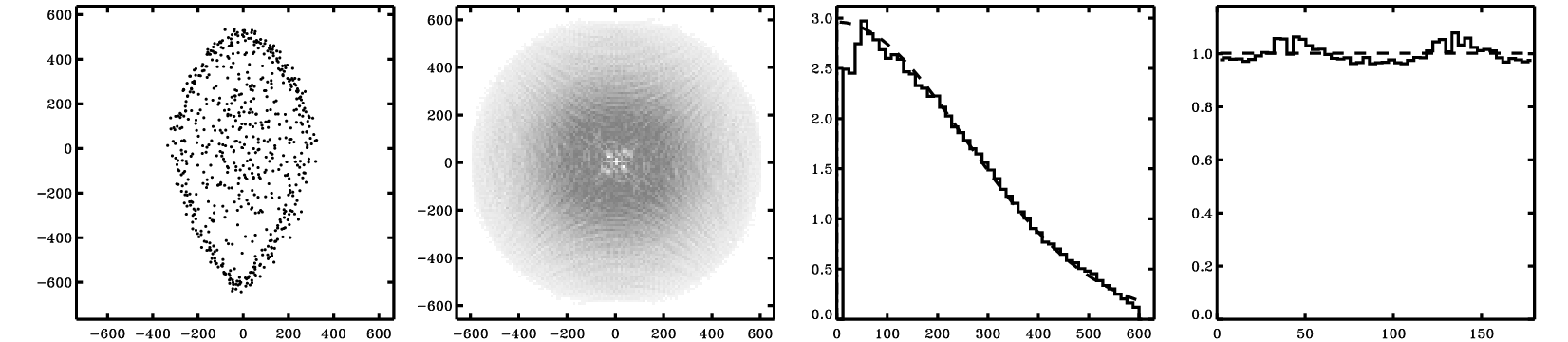}}}
\resizebox{17cm}{!}{\rotatebox{90}{\includegraphics{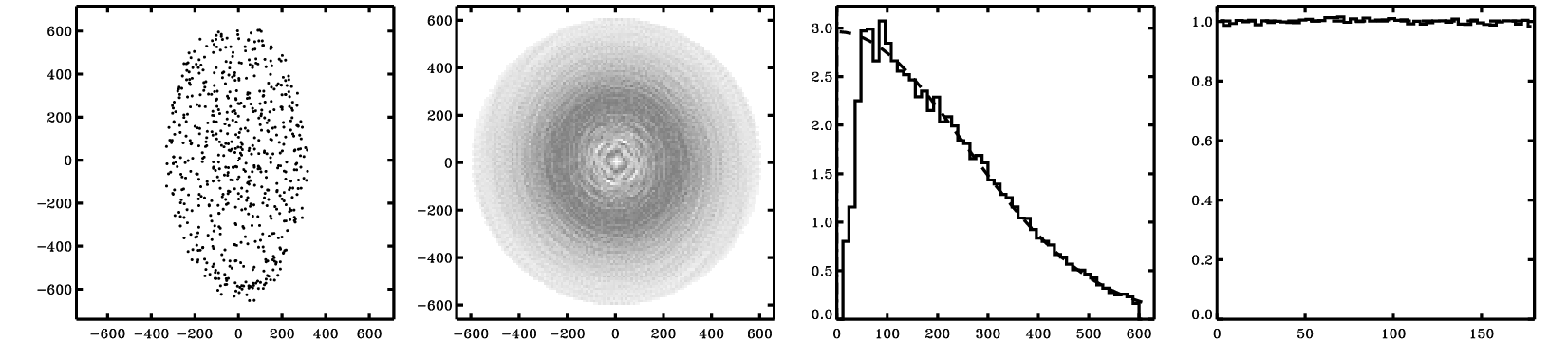}}}
\resizebox{17cm}{!}{\rotatebox{90}{\includegraphics{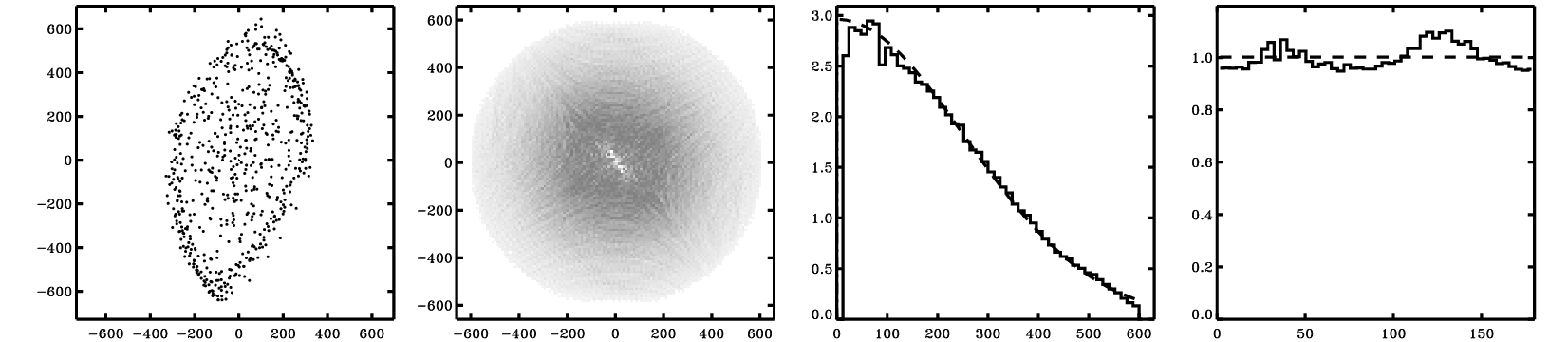}}}
     {\caption[]{Illustration of the configuration dependency on the declination of the source when observing with earth rotation synthesis. In 4 different situations 10 configurations of 64 antennas have been optimized independently. The antenna positions of the 10 configurations are plotted on the first figure then the averaged 2D-density and its radial and azimuthal profiles. In the profiles the density is normalized by the density of a uniform distribution of samples in the uv-disc of radius 600m$/\lambda$ ($\sigma_0=n_{visi}/\pi R^2$). On first 2 rows the source is observed from -1h to +1h and on the last two rows from -2h to 0h. For first and third rows the source is at $-83\deg$ of declination and for second and last rows the source is at $+37\deg$ of declination. The site latitude is $-23\deg$.}\label{fig:synth}}%
\end{figure*}
To illustrate the efficiency of the method an example is shown in Fig.\ref{fig:gaussian}: a randomly picked configuration of 64 antennas is optimized for a Gaussian distribution of FWHM, $\delta=0.7\times R$, $R$ being the radius of the sampled uv-disc. Fig.\ref{fig:conv} shows the convergence and quality of the optimization. The standard deviation as shown on the left-hand side was computed in the same kind of grids as those used to compute the gradients. The rapid and almost monotonic decrease of the standard deviation along iterations bring to evidence the rapid convergence. On the right-hand side the distribution of antennas of the optimized configuration is close to the Gaussian expected, confirming that the configuration is optimal or at least close to it.

Figure \ref{applic} shows the resulting configurations of several optimizations. Each row corresponds to a different optimization and the profile of the model distribution is represented in dashed line. The result obtained for a uniform distribution of samples (first row) can be compared to the Reuleaux triangle obtained by Keto. It can be noticed that to one extent the shape of the configuration confirms the analyze developed in \nocite{1997ApJ...475..843K}{Keto} (1997): it is a disturbed curve of constant width. Though, it is difficult to distinguish whether it is closer to a Reuleaux triangle than to a ring. In addition, some antennas (here 5 of the 64) are distributed in the center to compensate for the lack of weight at intermediate baseline lengths. Figure \ref{multiconf} shows the result of an optimization for a multi-configuration observation. Such an observation allows to get a better sensitivity on the short baselines.

Figure \ref{fig:synth} illustrates the dependency of the configuration on the declination of the source when observing with earth rotation synthesis. To get rid of initial conditions dependency and to emphasize the general tendency of the configurations, 10 configurations have been optimized for each situation. The first two rows consider observations of a source for an hour angle interval symmetric with respect to the transit : [-1h,+1h]. The first one at $60\deg$ from zenith in south direction and second at  $60\deg$ from zenith in the north direction. It can be seen that for the northern source more antennas need to be at the edges of the configuration which has a slightly different shape. For asymmetric hour angle interval ([-2h,0h] in the last two rows) northern and southern source observations are also different. It can also be noticed that for a southern source symmetric or asymmetric hour interval does not make any difference in the configuration. This is due to the almost circular shape of the tracks. For the northern sources the tracks are open ellipses and the degree of freedom in arranging them is lower. But once found the optimal arrangement gives much better Gaussian than for circles in the central region, i.e. for short baselines (see the profiles in Fig.\ref{fig:synth}). These examples show that the shape of the tracks has a strong impact on the configuration and that it is difficult to anticipate from a configuration optimized for a snapshot observation the way it should be modified for synthesis. This justifies the introduction of synthesis in the optimization program.

The efficiency and the flexibility of the method as illustrated by these examples make APO a well adapted tool for the array design. Interferometers have generally more stations than antennas to offer a range of resolutions and  multi-configuration observations. A procedure to optimize the locations of all the stations and solve the design problem as introduced Sect.\ref{sec:intro} is proposed:
\begin{enumerate}

\item Define a set of $N_{S}$ scientific goals, e.g.  at 100GHz imaging with a resolution of $1\arcsec$, 0.5$\arcsec$, 0.1$\arcsec$ and astrometry with a resolution of  0.1mas ($N_{S}$=4). To each of these goals corresponds a model distribution for the Fourier samples and a duration of observation to be used in the optimization (see Paper II for the case of imaging). 
 
\item Define a set of $N_{O}$ observational situations (the durations of observation are already determined from step 1), e.g. 3 representative declinations: at zenith, at $60\deg$ toward the South and at $60\deg$ toward the North, and only symmetric observations with respect to the transit ($N_{O}$=3). Note that if computing time is not a limitation instead of taking a set of representative declinations it is possible to take a set of subsets made of several declinations covering an interval and bearing  weights related to the distribution of sources over this interval.

\item Optimize $N_{S}\times N_{O}$ configurations corresponding to the possible combinations going from the most compact to the most extended and taking into account the terrain constraints. At each optimization try to use as much pads as possible of the previously optimized configurations.

\item If the total number of pads is too large merge some of them and do last step again or change the initial set of scientific purposes and start from step 1 again.
\end{enumerate}
This scheme is currently used for ALMA design \nocite{boone.m}({Boone} 2001a) in parallel with other approaches. It allows the design to be dictated by the scientific purposes and not by any {\it a priori} on the shapes of the configurations. It therefore warranties an optimal scientific return for the financial and technical effort invested in the project. 
\section{Conclusion}\label{sec:conclu}

A new method was introduced to solve the configuration problem. It is based on the computation of pressure forces emanating from the discrepancies between the model and actual distribution of Fourier samples. The array elements submitted to these forces move toward optimal positions. The efficiency, rapidity and flexibility of the method was demonstrated. Its implementation in a C++ library able to handle most of the observational situations like synthesis, mosaicing, multi-configuration as well as terrain constraints constitutes a powerful tool for array design. A procedure to manage multiple scientific goals and source positions was given in that purpose. 

Such an approach centered on the scientific purposes and able to handle a large number of antennas can contribute to the design of new generation interferometers like ALMA (64 antennas and 256 pads) or ATA (350 antennas). In the case of ALMA, even after the construction of the pads the degree of freedom in positioning the antennas for an observation will be high and it might be useful to have a software able find out which pads should be optimally used for a single observation and a single scientific purpose. APO could also achieve this task.
\begin{acknowledgements}
I am grateful to the referee T. Cornwell for recommendations that helped to considerably improve the clarity of this paper.
I thank J. Lequeux and A. L. Melchior for careful reading of the manuscript and encouraging comments.
\end{acknowledgements}


\end{document}